\documentstyle[aps,prl,preprint]{revtex}
\begin{document}
\draft
\title{Quantum Information Processing with Semiconductor Macroatoms}
\author{Eliana Biolatti$^{1,2}$, Rita C. Iotti$^{1,2}$, 
Paolo Zanardi$^{1,3}$, and Fausto Rossi$^{1,2,3}$}
\address{$^1$ Istituto Nazionale per la Fisica della Materia (INFM)}
\address{$^2$ Dipartimento di Fisica, Politecnico di Torino,
Corso Duca degli Abruzzi 24, I-10129 Torino, Italy}
\address{$^3$ Institute for Scientific Interchange (ISI) foundation, 
Villa Gualino, Viale Settimio Severo 65, I-10133 Torino, Italy}
\date{\today}
\maketitle
\begin{abstract}

An {\it all optical} implementation of quantum information processing with 
semiconductor macroatoms is proposed. Our quantum hardware consists of an 
array of semiconductor quantum dots and the computational degrees of freedom 
are energy-selected interband optical transitions. 
The proposed quantum-computing strategy exploits exciton-exciton interactions 
driven by ultrafast sequences of multi-color laser pulses.
Contrary to existing proposals based on charge excitations, the present 
all-optical implementation does not require the application of 
time-dependent electric fields, thus allowing for a sub-picosecond, 
i.e. decoherence-free, operation time-scale in realistic state-of-the-art 
semiconductor nanostructures.

\end{abstract}
\narrowtext

The introduction of quantum information/computation (QIC) \cite{QIC} 
as an abstract concept,
has given birth to a great deal of new thinking about how to design 
and realize quantum information processing devices.
This goal  is extremely challenging:
one should be able to perform, on a system with a well-defined 
quantum state space (the {\it computational} space), 
precise quantum-state preparation, 
coherent quantum manipulations ({\it gating}) of arbitrary length,
and state detection  as well.
It is well known that the major obstacle to implement this ideal
scheme is {\it decoherence}: the spoiling of the unitary character 
of quantum evolution
due to the uncontrollable coupling with environmental, 
i.e., non-computational, degrees of freedom.
Mostly due to the need of low decoherence  rates, 
the first proposals for experimental realizations of quantum information  
processing devices originated from specialties in atomic
physics \cite{AP}, in quantum optics \cite{QO}, and in 
nuclear and electron magnetic-resonance spectroscopy \cite{NMR-EMR}. 
On the other hand, practically relevant quantum computations require
a large number of quantum-hardware units ({\it qubits}), that
are known to be hardly  achievable in terms of such  systems. 
In contrast, in spite of the serious difficulties related to the ``fast'' 
decoherence times, a solid-state implementation of QIC
 seems to be the only way to benefit synergistically from the 
recent progress in ultrafast optoelectronics \cite{UO} as well as in 
meso/nanostructure fabrication and characterization \cite{QD}.
Among the proposed solid-state implementations one should 
mention those in superconducting device physics \cite{SC} and in meso- and 
nanoscopic physics \cite{QD-spin}. In particular, the first 
semiconductor-based proposal, by Loss and DiVincenzo, relies on spin 
dynamics in quantum dots; it exploits the low decoherence of
spin degrees of freedom  in comparison to the one of charge excitations. 

As originally envisioned in \onlinecite{ZR}, 
gating of charge excitations could be performed by exploiting  {\it present} 
 ultrafast laser technology \cite{UO}, that
  allows  to generate and manipulate electron-hole
quantum states on a sub-picosecond time-scale:
{\em  coherent-carrier-control} \cite{CCC}.
In this respect, decoherence times on nano/microsecond scales 
can be regarded as ``long'' ones.
Based on this idea a few implementations 
have been recently put forward \cite{SBI};  
However, while in these 
proposals single-qubit operations are implemented  by means  of ultrafast 
optical spectroscopy,  the control of two-qubit 
operations still involves the application of external fields 
and/or microcavity-mode couplings,
whose switching times are much longer than
decoherence times in semiconductor quantum dots (QDs). 
It clearly follows  that such proposals are currently out of reach in terms
of state-of-the-art optoelectronics technology.

As already pointed out in \cite{ZR}, in order to take full advantage 
from modern ultrafast laser spectroscopy 
one should be able to design fully optical gating schemes
able  to perform single- {\it and}  two-qubit operations on a 
sub-picosecond time-scale.
Following this spirit, in this Letter we propose the first {\it all-optical} 
implementation with semiconductor macromolecules. 
Our analysis is based on a realistic,  fully 
three-dimensional, description of multi-QD structures, whose many-body 
electron-hole Hamiltonian can be schematically written as \cite{SST}:
\begin{equation}
{\bf H} = {\bf H}_\circ + {\bf H}' = \left({\bf H}_c + {\bf H}_{cc}\right) + 
\left({\bf H}_{cl} + {\bf H}_{env}\right)\ ,
\label{H}
\end{equation} 
where 
${\bf H}_c$ 
describes the non-interacting electron-hole system within the 
nanostructure confinement potential,
${\bf H}_{cc}$
is the sum of the three (electron-electron, hole-hole, and electron-hole) 
Coulomb-interaction terms, 
${\bf H}_{cl}$
describes the coupling of the carrier system with a classical light 
field \cite{cl}, while ${\bf H}_{env}$ describes the interaction of the carrier 
system with environmental degrees of freedom, like phonon and plasmon 
modes of the host material.
The latter is responsible for decoherence processes and it will be 
accounted for within a density-matrix formalism (see below).

For any given number of electron-hole 
pairs $N$, a direct 
diagonalization of 
${\bf H}_\circ$ 
will provide the many-body states of the interacting electron-hole system; 
they, in turn, allow to evaluate many-exciton 
optical spectra, i.e.,  the absorption probability 
corresponding to the generic $N \to N'$ transition.

The above theoretical scheme has been applied to realistic state-of-the-art
QD arrays. In particular, as quantum hardware, we consider a GaAs-based 
structure with in-plane parabolic confinement potential \cite{2D-pp}; 
Moreover, as discussed below (see Fig.~\ref{fig2}), in order to induce a 
significant exciton-exciton dipole coupling, 
an in-plane static 
electric field $F$ is applied. 
The square-like carrier confinement along the growth ($z$) direction for 
electrons and holes is schematically depicted in Fig.~\ref{fig1} for the 
array unit cell $a+b$. 
We stress that the geometrical and material parameters of the proposed 
prototypical structure in Fig.~\ref{fig1} are fully compatible with current
QD growth and characterization technology \cite{QD,CCC}.

Let us discuss first the optical response of the 
{\it semiconductor macromolecule} ($a+b$) in Fig.~\ref{fig1}. 
The excitonic ($0 \to 1$) optical spectrum in the presence of an in-plane 
electric field $F = 15$\,kV/cm is shown in 
Fig.~\ref{fig2}(A), where the two lowest optical transitions 
correspond to the formation of direct ground-state excitons in dot $a$ and 
$b$, respectively (see Fig.~\ref{fig1}). 
In contrast, the high-energy peaks correspond to optical transitions 
involving excited states of the in-plane parabolic potential.
Due to the strong in-plane carrier confinement, the low-energy excitonic 
states are expected to closely resemble the 
corresponding single-particle ones, 
thus involving the parabolic-potential ground state only.
This is confirmed by the excitonic spectrum (solid curve) in 
Fig.~\ref{fig2}(B), which has 
been obtained limiting our single-particle basis set to the 
parabolic-potential ground state. As we can see, apart from a small rigid
shift, the relative position of the lowest transitions is 
the same. 
This suggests to use as a basis of our computational space the set 
formed by the lowest excitonic transition in each QD.

Let us now come to the biexcitonic spectrum [dashed line in 
Fig.~\ref{fig2}(B)]; 
it describes the generation of a second electron-hole 
pair in the presence of a previously created exciton ($1 \to 2$ 
optical transitions). 
The crucial feature in Fig.~\ref{fig2}(B) is the magnitude of the "biexcitonic 
shift" \cite{QD},   
i.e., the energy difference between the excitonic and the biexcitonic 
transition (see solid and dashed curves).
For the QD structure under investigation we get energy splittings up 
to $5$\,meV [see inset in Fig.~\ref{fig2}(B)], 
which is by far  larger than typical biexcitonic 
splittings in single QDs \cite{QD}. 
This is due to the in-plane static field $F$, 
which induces a charge separation between electrons and holes. 
This, in turn, gives rise to significant dipole-dipole
coupling between adjacent excitonic states. 
The microscopic origin of such exciton-exciton coupling is 
the same of the Forster process exploited by Quiroga and Johnson \cite{Q-J}
for the generation of entangled states in coupled QDs.

The physical origin of the biexcitonic shift $\Delta {\cal E}$ 
is qualitatively described in
Fig.~\ref{fig2}(C), where we show the electron and hole charge distribution 
corresponding to the biexcitonic ground state.

The central idea in our QIC proposal is to exploit such few-exciton effect 
to design {\it  conditional operations}. 
To this end let us introduce the excitonic occupation number operators 
$\hat n_l$,
where $l$ denotes the generic QD in our array.
The two states with eigenvalues $n_l=0$ and $n_l=1$ correspond, 
respectively, to the absence (no conduction-band electrons) and to the 
presence of a ground-state exciton (a Coulomb-correlated electron-hole pair) 
in dot $l$;
they constitute our single-qubit 
basis: $\vert 0 \rangle_l$ and 
$\vert 1 \rangle_l$.
The whole  computational state-space ${\cal H}$ is then spanned
by the basis 
$\vert{\bf{n}}\rangle = 
\otimes_{l} \vert n_l \rangle,\,(n_l=0,\,1)$.

The full many-body Hamiltonian ${\bf H}_\circ$ in (\ref{H}) restricted to the 
above computational space ${\cal H}$ reduces to
\begin{equation}
\tilde {\bf{H}}_\circ = \sum_{l} {\cal E}_l\,\hat n_l + 
{1\over 2} \sum_{ll'} \Delta{\cal E}_{ll'}\,\hat n_l\,\hat n_{l'} \ .
\label{H0tilde} 
\end{equation}
Here, ${\cal E}_l$ denotes the energy of the ground-state exciton in dot $l$ 
while $\Delta{\cal E}_{ll'}$ is the biexcitonic shift due to the Coulomb 
interaction between dots $l$ and $l'$, previously introduced 
(see Fig.~\ref{fig2}). 
The effective Hamiltonian in (\ref{H0tilde}) has exactly the same structure 
of the one proposed by Lloyd in his pioneering paper on quantum cellular 
automata \cite{QCA}, and it is the Model Hamiltonian currently used
in most of the NMR quantum-computing schemes \cite{Cory}. 
This fact is extremely important since it tells us that: 
(i) the present semiconductor-based implementation contains all 
relevant ingredients for the realization of basic QIC processing;
(ii) it allows to establish a one-to-one correspondence between 
our semiconductor-based scheme and much more mature implementations, like
NMR \cite{Cory}.

According to (\ref{H0tilde}),
the single-exciton  energy ${\cal E}_l$ is renormalized 
by the biexcitonic shift $\Delta{\cal E}_{ll'}$, induced by the presence of
a second exciton in dot $l'$ ($\tilde{\cal E}_l = {\cal E}_l + 
\sum_{l' \ne l} \Delta{\cal E}_{ll'}\, n_{l'}$).
In order to better illustrate this  idea, 
let us focus again on the two-QD structure, 
i.e. two-qubit system, of Fig.~\ref{fig1}
and fix our attention on one of the two dots, say dot $b$.   
The effective energy gap between  $\vert 0 \rangle_b$ and $\vert 1 \rangle_b$ 
depends now on the occupation of dot $a.$
This elementary remark suggests to design properly tailored laser-pulse 
sequences to implement 
controlled-not logic gates among the two QDs as well as
single-qubit rotations.
Indeed, by sending an ultrafast laser $\pi$-pulse with central 
energy 
$\hbar\omega_b[n_a] = {\cal E}_b + \Delta{\cal E}_{ba} n_a$, 
the transition $\vert n_b \rangle_b \rightarrow \vert 1-n_b \rangle_b$ 
($\pi$-rotation)
of the {\em target} qubit (dot $b$) is obtained if and only if the 
{\em control} qubit (dot $a$) is in the state $\vert n \rangle_a$. 
Notice that the above  scheme corresponds to the 
so-called selective population transfer in NMR \protect\cite{Cory}\protect; 
alternative procedures used in that field can be
adapted to the present proposal as well. 

Moreover,
by denoting with $U_b^{n_a}$ the generic unitary transformation induced by
the laser $\pi$-pulse of central frequency $\omega_b[n_a]$, 
it is easy to check that the two-color pulse sequence 
$U_b^0\,U_b^1$ 
achieves the unconditional $\pi$-rotation of qubit $b$.

In order to test the viability of the proposed quantum-computation 
strategy,
we have performed a few simulated experiments of basic quantum information 
processing. 
Our time-dependent simulations are based on the realistic state-of-the-art QD
structure of Fig.~\protect\ref{fig1}\protect: 
${\cal E}_a = 1.70$\,eV,
${\cal E}_b = 1.71$\,eV,
$\Delta{\cal E} = 4,5$\,meV, which correspond to $F=30$\,kV/cm   
[see inset in Fig.~\protect\ref{fig2}(B)\protect].
They are based on a  numerical solution of the 
Liouville-von Neumann equation
describing the exact quantum-mechanical evolution of the many-exciton 
system (\ref{H0tilde}) within our computational subspace ${\cal H}$ in the 
presence of environment-induced decoherence processes 
[see term ${\bf H}_{env}$ in Eq.~(\ref{H})] \cite{T1-T2}. 
Figure \ref{fig3} shows a simulated sequence of single- plus two-qubit 
operations driven by a two-color laser-pulse sequence.

Initially
the system is in the state $\vert 0,0 \rangle \equiv 
\vert 0 \rangle_a \otimes \vert 0 \rangle_b$. 
The first laser pulse 
(at $t = 0.2$\,ps) is tailored 
in such a way to induce a $\pi\over 2$ rotation of the qubit 
$a$: $\vert 0,0 \rangle \to (\vert 0,0 \rangle + \vert 1,0 \rangle)/\sqrt{2}$. 
At time $t = 0.8$\,ps a second pulse induces a conditional $\pi$-rotation of 
the qubit $b$:
$\vert 0,0 \rangle + \vert 1,0 \rangle \to 
\vert 0,0 \rangle + \vert 1,1 \rangle$.
This last operation plays a central role in any quantum information 
processing, since it transforms a {\it factorized} state 
($(\vert 0 \rangle + \vert 1 \rangle) \otimes \vert 0 \rangle$) 
into an  {\it entangled} state.
The scenario described so far is confirmed by the time evolution of the 
exciton occupation numbers $n_a$ and $n_b$ reported in (A) 
as well as of the diagonal elements of the density matrix
(in our four-dimensional computational basis) reported in (B).
As we can see, during the pulse energy-nonconserving (or off-resonant) 
transitions \cite{SST} take place; however, at the end of the pulse 
such effects vanish and the desired quantum state is reached.

The experiment simulated above clearly shows that the energy scale of the 
biexcitonic splitting in our quantum-dot molecule [see Fig.~\ref{fig2}] is 
compatible with the sub-picosecond operation time-scale of Fig.~\ref{fig3}.

At this point a few comments are in order.
First we stress a very important feature of the proposed semiconductor-based 
implementation:
as for NMR quantum computing, 
two-body interactions 
are always switched on 
(this  should be compared to the schemes in which 
two-qubit gates are realized by turning on and off the coupling between 
subsystems, e.g., by means of slowly-varying fields and cavity-mode 
couplings);
conditional as well as unconditional dynamics
is realized by means of sequences of ultrafast  single-qubit operations 
whose length 
does not scale as a function of the total number of QDs in the array 
\cite{NN}.

Let us now come to the {\it state measurement}. 
In view of the few-exciton character of the proposed quantum hardware,
the conventional measurement of the carrier subsystem 
by spectrally-resolved luminescence needs to be replaced by more 
sensitive detection schemes. 
To this end, a viable strategy could be to apply to our semiconductor-based
structure the well-known recycling techniques commonly used in 
quantum-optics experiments \cite{recycling}.
Generally speaking, the idea is to properly combine quantum- and 
dielectric-confinement effects in order to obtain well-defined energy levels, 
on which to design energy-selective photon-amplification schemes. 

The nanoscale range of the inter-dot coupling we employed for 
enabling conditional dynamics does not allow for space-selective
optical addressing of individual qubits. 
For this reason, at least for our basic QD molecule ($a + b$), we resorted to 
an energy selective addressing scheme. 
However, extending such strategy to the whole QD array would imply 
different values of the excitonic transition in each QD, 
i.e., ${\cal E}_l\neq {\cal E}_{l'}$.  
This, besides obvious technological difficulties, would constitute a conceptual
limitation  of  scalability  towards  massive Quantum Computations.
The problem can be avoided following a completely different strategy 
originally proposed by Lloyd \cite{QCA} and recently improved in 
\onlinecite{Benjamin}: 
by properly designed sequences of multicolor global pulses within a 
cellular-automaton scheme, local addressing is replaced by 
information-encoding transfer along our QD array.

Finally, a present limitation of the proposed quantum hardware are the 
non-uniform structural and geometrical properties of the QDs in the array, 
which may give rise to energy broadenings larger than the biexcitonic 
shift. 
However, recent progress in QD fabrication
---including the realization of QD structures in microcavities--- 
will allow, we believe, to overcome this purely technological (non conceptual) 
limitation.

In summary, the first {\it all optical implementation of QIC with 
a semiconductor-based quantum hardware} has been proposed. Our analysis has
shown that energy-selected optical transitions in realistic state-of-the-art 
QD structures are good candidates for quantum-information encoding and 
manipulation. The sub-picosecond time-scale of ultrafast laser spectroscopy
allows for a relatively large number of elementary operations within the 
exciton decoherence time.

\medskip\par
We are grateful to David DiVincenzo, Neil Johnson, Seth Lloyd, and Mario 
Rasetti for stimulating and fruitful discussions.

This work has been supported in part by the European Commission through the
Research Project {\it SQID} within the {\it Future and Emerging 
Technologies (FET)} programme.

\begin{figure} 
\caption{Schematic representation of the square-like potential profile for
electrons (e) and holes (h) along the growth ($z$) direction of our QD array.
This is tailored in such a way to allow for
an energy-selective creation/destruction of bound electron-hole pairs (i.e.,
excitons) in dots $a$ and $b$.
Moreover, the inter-dot barrier width ($w$ $\sim$ $50$ \AA ) is such to prevent
single-particle tunneling and at the same
time to allow for significant inter-dot Coulomb coupling.}
\label{fig1}
\end{figure}
\begin{figure} 
\caption{Optical response of the array unit cell ($a+b$) in 
Fig.~\protect\ref{fig1}.
(A) Excitonic spectrum obtained including the realistic multilevel structure 
of the in-plane parabolic potential. 
(B) Excitonic (solid curve) and biexcitonic spectrum (dashed curve) 
obtained including the in-plane ground state only. 
Due to the well-defined polarization of our laser source, 
the two structures in the biexcitonic spectrum correspond to the 
formation of an exciton in dot $a$ given an exciton in dot $b$ 
and {\it vice versa}.
(C) Three-dimensional view of the spatial charge distributions of the 
two electrons ($e_a$ and $e_b$) and holes ($h_a$ and $h_b$) 
corresponding to the biexcitonic ground state in (B).
As we can see, the charge separation
induced by the static field increases significantly the average distance
between electrons and holes, thus decreasing their attractive interaction.
On the other hand, the repulsive terms are basically field independent.
This is the origin of the positive energy difference  $\Delta {\cal E} $  in
(B).     
} 
\label{fig2}
\end{figure}
\begin{figure} 
 \caption{Time-dependent simulation of single (unconditional) plus 
two-qubit (conditional) operations (see text).
(A) Exciton populations ($n_a$ and $n_b$), and (B) 
diagonal elements of the density matrix as a function of time. The two-color 
pulse sequence is also sketched schematically.}
\label{fig3}
\end{figure}

\begin{references}
\bibitem{cl}
The carrier-light (cl) coupling  term is responsible for the 
creation/recombination of electron-hole pairs and it will 
induce the desired single-qubit rotations.
\bibitem{2D-pp} The interlevel energy splitting of the in-plane parabolic 
potential has been chosen to be $30$\,meV for electrons and $16$\,meV for 
holes.
\bibitem{T1-T2}
Decoherence processes are accounted for in our density-matrix formalism 
by means of a 
standard $T_1 T_2$ model: we employ a band-to-band recombination 
time $T_1 = 1$\,ns and we describe phonon-induced decoherence processes in 
terms of an exciton-phonon dephasing time $T_2 = 50$\,ps, 
which is compatible with the experimental values given in \onlinecite{CCC}.
\bibitem{NN}
Due to the short-range nature of the exciton-exciton 
dipole interaction, we are allowed to consider nearest-neighbor coupling only:
indeed, for the specific quantum hardware considered (see Fig.~\ref{fig1}),
in view of the ${1\over r^3}$ behaviour and, more important, of their 
small oscillator strengths, second-neighbor couplings play no significant 
role.
\end{references}
\end{document}